\documentclass[12pt]{article}
\usepackage{hyperref}
\usepackage{amssymb,latexsym,amsmath} 
\usepackage{authblk}
\usepackage[utf8]{inputenc}
\usepackage{dsfont}
\usepackage[utf8]{inputenc}
\usepackage{graphics}
\usepackage{color}
\usepackage{subfig}
\usepackage{graphicx}
\usepackage{mathrsfs}
\begin{document}
\title{\bf Callan-Symanzik-like equation in information theory}

\author{Mojtaba Shahbazi \thanks{mojtaba.shahbazi@modares.ac.ir}, Mehdi Sadeghi\thanks{mehdi.sadeghi@abru.ac.ir}\,\,\, \hspace{2mm}\\
	{\small {\em Department of Physics, Faculty of Basic Sciences,}}\\
	{\small {\em Ayatollah Boroujerdi University, Boroujerd, Iran}}
}
\date{\today}
\maketitle

\abstract{Within the "complexity=anything" proposal of holography, the complexity growth rate (CGR) can exhibit jumps, interpreted as phase transitions. We demonstrate that the location and amplitude of these jumps are governed by the dynamics of bulk fields, which, via the fluid-gravity correspondence, map to the boundary energy-momentum tensor. The behavior of the CGR near these critical points exhibits scaling and universality. We show that the CGR satisfies a Callan-Symanzik-like equation near the transitions. Our results provide a new information-theoretic interpretation of the Callan–Symanzik equation, with the CGR running with the energy scale.}

\section{Introduction}
The AdS/CFT correspondence establishes a profound duality between a gravitational theory in anti-de Sitter (AdS) space and a conformal field theory (CFT) on its boundary \cite{malda}. This holographic framework provides a powerful dictionary for studying strongly-coupled quantum systems via their dual classical gravitational descriptions. A prominent application lies in black hole physics, where the eternal growth of the Einstein-Rosen bridge volume in the bulk, for a system in thermal equilibrium, posed a conceptual puzzle. Recent insights suggest this geometric evolution corresponds to the computational complexity of the boundary quantum state, the minimum number of fundamental operations required to prepare the target state from a reference state \cite{3lec}.

Several holographic conjectures formalize this connection. The most studied are "complexity=volume" (CV), which identifies complexity with the volume of maximal spatial slices \cite{cv}, "complexity=action" (CA), which equates it with the gravitational action on the Wheeler-DeWitt patch \cite{ca}, and "complexity=spacetime volume" (CV2.0) \cite{cv2}. More recently, a generalized framework, "complexity=anything", $\mathcal{C}_{Any}$ has been proposed \cite{general}. This conjecture posits that complexity can be dual to a vast class of diffeomorphism-invariant observables constructed from arbitrary scalar functionals of the bulk metric. A fascinating feature of the $\mathcal{C}_{Any}$ proposal is that the complexity growth rate (CGR) can develop multiple branches and exhibit sudden jumps as a function of time, which have been interpreted as signatures of computational phase transitions in the boundary theory \cite{phase,phase1}. This work investigates the physical origin of these transitions. We address two central questions:
\begin{itemize}
\item What boundary physics determines the amplitude of the jumps in the complexity growth rate?\\
\item What physics controls their temporal location?
\end{itemize}
We find that the jump locations are dynamically set by bulk fields. Through the fluid-gravity duality, these map to the boundary energy-momentum tensor \cite{sken,kss}. The jump amplitudes, conversely, are linked to the renormalization group (RG) flow of the boundary theory.

Remarkably, the CGR near these critical points exhibits scaling behavior and universality. We show that it satisfies a Callan-Symanzik-like (CSL) equation, where the beta function is expressed in terms of the generalization parameter from the $\mathcal{C}_{Any}$, proposal. This criticality displays universal features, independent of the specific choice of the generalized functional.

Very recently, a detailed analysis of phase transitions within the "complexity = anything'' framework has appeared in \cite{2503}. That work studies the appearance of multiple extremal surfaces in two-dimensional theories of dilaton gravity and their interpretation in terms of circuit complexity, offering a microscopic perspective on the competition between branches. While the setup there differs in some details, the underlying phenomenon of branch competition and critical transitions is closely related to the CSL behavior we uncover here. We will comment further on this connection in the conclusion.

The paper is organized as follows. In the section \ref{sec2}, we review the $\mathcal{C}_{Any}$ framework and derive the CGR for a generic bulk metric. The section \ref{control} establishes the connection between the jump locations and the dynamics of boundary field theory quantities. In the section \ref{cr}, we analyze the critical scaling of the CGR, derive the associated critical exponents, and present the central result: the CSL equation in the context of information theory. We conclude in the section \ref{con} with a discussion of the implications for quantum information processing and future directions.
\section{Complexity growth rate in the $\mathcal{C}_{Any}$ proposal}
\label{sec2}
Recently, $\mathcal{C}_{Any}$ (complexity equals anything) conjecture has emerged as a generalized version of complexity in the bulk. In this proposal the complexity is not limited to a specific geometric quantity but can be described by a more general function to include a class of new diffeomorphism invariant observables \cite{general}:
\begin{align}\label{obser}
	\mathcal{O}_{F_1,\Sigma_{F_2}}(\Sigma_{CFT})=\frac{1}{G_N L}\int_{\Sigma_{F_2}}\mathrm{d}^{d}\sigma \sqrt{h} F_1(g_{\mu\nu};X^{\mu})
\end{align}
where $F_1$ can be a general scalar function of metric $g_{\mu\nu}$  and an embedding $X^{\mu}(\sigma^a)$ of the hypersurfaces. Furthermore, $\Sigma_{F_2}$ is codimension-one hypersurface in the bulk spacetime with boundary time slice $\partial\Sigma_{F_2} =\Sigma_{CFT}$. Extremality of the hypersurface leads to:

\begin{align}\label{var}
	\delta_{X}\Big(\int_{\Sigma}\mathrm{d}^{d}\sigma \sqrt{h} F_2(g_{\mu\nu};X^{\mu})\Big)=0. 
\end{align}
For simplicity, we follow the case $F_1 =F_2$, so the observable \eqref{obser}, known by $\mathcal{C}_{Any}$, obeying above condition is expressed by:

\begin{align}\label{maxv}
	\mathcal{C}_{Any}(\tau)= \max_{\partial\Sigma(\tau)=\Sigma_{\tau}}\frac{V_x}{G_N L}\left[\int_{\Sigma}\mathrm{d}^{d}\sigma \sqrt{h} F_1(g_{\mu\nu};X^{\mu})\right],
\end{align}
where $h$ is the determinant of induced metric on the given hypersurface.
One of the usual choice of generalization is:

\begin{align}\label{F1F2}
	F_1=F_2=1+\gamma L^4 \mathcal{C}^2,
\end{align}
where $\mathcal{C}^2$ is the Weyl tensor squared and for $\gamma=0$ the standard CV proposal is recovered. In the following, we are going to study the $\mathcal{C}_{Any}$ of a generic metric:

\begin{align}\label{metricEF}
	\mathrm{d} s^{2}=g_{tt} \mathrm{d} t^{2}+ g_{rr}\mathrm{~d}r^2+ \frac{r^2}{L^2} d\Omega^2.
\end{align}
To do this, we need to transform the metric to Eddington-Finkelstein coordinates:

\begin{align}\label{metricEF}
	\mathrm{d} s^{2}=g_{tt} \mathrm{d} v^{2}- 2\sqrt{-g_{tt}g_{rr}} \mathrm{~d} v \mathrm{~d} r+ \frac{r^2}{L^2}d\Omega^2,
\end{align}
where:

\begin{align}
	v=t+r^{*}(r), \quad \mathrm{d}r^{*}=\sqrt{-\frac{g_{rr}}{g_{tt}}}\mathrm{d}r.
\end{align}
Then, the complexity could be computed from \eqref{maxv} and \eqref{F1F2} as \cite{general}:

\begin{align}\label{Any}
	\mathcal{C}_{Any}(\tau)= \frac{V_0}{G_N L}\int_{\Sigma}\mathrm{d}\sigma r^{d-1}\sqrt{g_{tt} \dot{v}^{2}-2 \sqrt{-g_{tt}g_{rr}}\dot{v} \dot{r}} a(r),
\end{align}
where $a=F_1$, the dots indicate derivatives with respect to $\sigma$, and $V_0$ represents the volume of spatial directions.

\begin{figure}[!h]
	\centering
	\includegraphics[width=9cm]{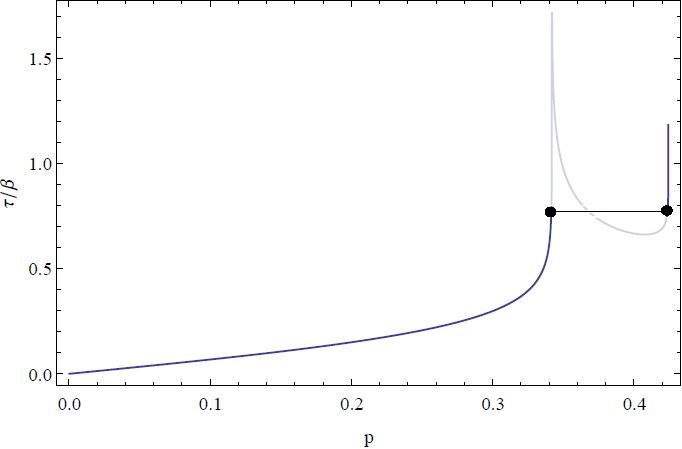} 
	\caption{a schematic behavior of CGR in terms of the boundary time. The black line shows the phase transition where the complexity has an intersection in terms of $r_{min}$ \cite{phase}.}\label{cgr}
\end{figure}

Supposing $\mathcal{C}_{Any}$ as an action leads to a conserved  momentum conjugate to coordinate $v$, because of spacetime is stationary:
 
\begin{align}
	P_v=-\frac{\partial \mathcal{L}}{\partial \dot{v}}=-\frac{a(r) r^{d-1}(g_{tt} \dot{v}- \sqrt{-g_{tt}g_{rr}}\dot{r})}{\sqrt{g_{tt} \dot{v}^{2}-2 \sqrt{-g_{tt}g_{rr}}\dot{v} \dot{r}}}=\frac{r^{d-1}}{H(r)}\Big(\dot{r}-\sqrt{-\frac{g_{tt}}{g_{rr}}} \dot{v}\Big).	
\end{align}
Due to the fact that \eqref{Any} is diffeomorphism invariant and doesn't change under reparametrization, the parameter $\sigma$ could be fixed by choosing:

\begin{align}
	\sqrt{g_{tt} \dot{v}^{2}-2 \sqrt{-g_{tt}g_{rr}}\dot{v} \dot{r}}=a(r) \sqrt{-g_{tt}g_{rr}}H(r),
\end{align}
where $H$ is an arbitrary function. Then it is straightforward to derive extremality conditions:

\begin{align}\label{rdot}
	\dot{r}=\pm \sqrt{\frac{H(r)^2 P^2_v}{r^{2(d-1)}} -g_{tt}a(r)^2H(r)^2},
\end{align}
\begin{align}\label{tau}
	\dot{t}=\dot{v}-\sqrt{-\frac{g_{rr}}{g_{tt}}}\dot{r}=\frac{-\sqrt{-\frac{g_{rr}}{g_{tt}}}P_v\dot{r}H(r)}{r^{d-1} \sqrt{\frac{H(r)^2P^2_v}{r^{2(d-1)}}- U(r)}},
\end{align}
where $U(r)$ is an effective potential.

\begin{align}
	U(r)=g_{tt}a(r)^2H(r)^2.
\end{align}
Then, \eqref{rdot} gives the equation of motion like for a classical particle:

\begin{align}\label{newu}
	\dot{r}^{2}+U(r)=\frac{H(r)^2P^2_v}{r^{2(d-1)}}.
\end{align}
In the symmetric trajectory, the conserved momentum is a function of turning point $r_{min}$, we could redefine the potential as follows\footnote{At $r_{min}$ the function $H(r)$ is redundant because it is omitted from both sides of \eqref{newu}, so we could redefine new potential which is independent of $H(r)$.}:

\begin{align}\label{poten}
	\tilde{U}(r)=g_{tt}a(r)^2 r^{2(d-1)}.
\end{align}
Then:
	\begin{align}
		P^2_v=	\tilde{U}(r_{min})=g_{tt}(r_{min})a(r_{min})^2 r_{min}^{2(d-1)}.
	\end{align}	
Boundary time derivation of $\mathcal{C}_{Any}$ leads to:

	\begin{align}
		\frac{\mathrm{d} \mathcal{C}_{Any}}{\mathrm{~d} \tau}=\frac{1}{2} \frac{\mathrm{d} \mathcal{C}_{Any}}{\mathrm{~d} \tau_{\mathrm{R}}}&=\frac{V_0}{G_N L} P_v\nonumber\\
		&=\frac{V_0}{G_NL} \sqrt{g_{tt}(r_{min})}a(r_{min}) r_{min}^{(d-1)}.
	\end{align}
A schematic behavior of the complexity growth rate is depicted in Fig.(\ref{cgr}).
\\	

Then, the growth rate of the generalized complexity can be studied by the behavior of the conserved momentum in full-time. The late-time behavior of the growth rate is determined by: 

	\begin{align}\label{tauinf}
		\lim _{\tau \rightarrow \infty} \frac{\mathrm{d} \mathcal{C}_{Any}}{\mathrm{~d} \tau}=\frac{V_0}{G_NL} \sqrt{g_{tt}(\tilde{r}_{\min})} a(\tilde{r}_{\min }) \tilde{r}_{min}^{(d-1)},
	\end{align}
where $\tilde{r}_{\min }$ is local maximum of the effective potential, and because at this radius time goes to infinity, it is known sometimes as $r_f:= \tilde{r}_{\min }$.

\section{What controls these jumps in CGR?}\label{control}
The "complexity = anything" proposal exhibits jumps in the complexity growth rate while the former proposals do not. This phenomenon raises the question that why there are these jumps in this proposal. Are they related to a physical phenomenon? In this section we answer to the questions affirmatively and seek the reason in the boundary theory of the bulk.

\begin{figure}[!h]
	\centering
	\includegraphics[width=9cm]{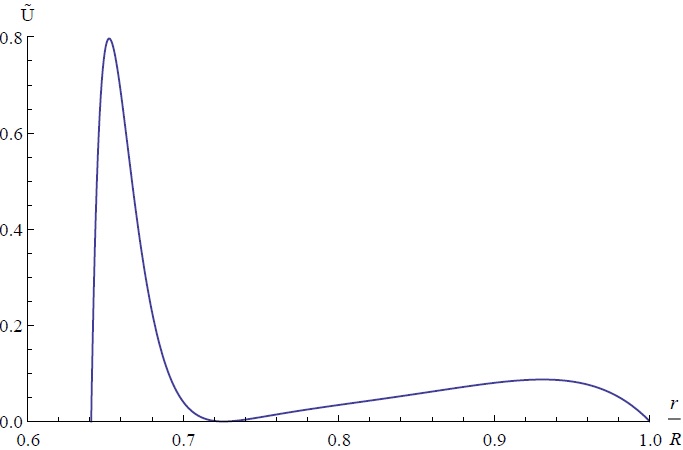} 
	\caption{A schematic behavior of the potential $\tilde{U}$ in terms of radial coordinate $r$ and the horizon $R$.}\label{u}
\end{figure}

The late time jump and consequently the sharp transition occurs around the asymptotic behavior of CGR, the extremum values of $P_v$, when the potential \eqref{poten} has a local maxima, in other words $\frac{d\tilde{U}}{dr}=0$ around the jumps (for details see Appendix \ref{potjump}). A schematic behavior of the potential $\tilde{U}$ is demonstrated in Fig.(\ref{u}).
\\	

Since the jump locations $r_i$ are determined by the behavior of $\tilde{U}(r)$, we can express this condition using the Fefferman-Graham expansion at finite $r_i$ cut-off, which directly links it to the boundary energy-momentum tensor and its properties. For a generic solution we have got:
\begin{align}
ds^2=&g_{tt}dt^2+g_{rr}dr^2+r^2d\Omega^2\\
\tilde{U}=&g_{tt} a^2 r^{2(d-1)}\\
&g_{tt}a^2r^{2(d-1)}=cte:=\alpha.\label{cte}
\end{align}
In the following we set $\alpha=1$. One could rewrite \eqref{cte} in Fefferman-Graham coordinates as following (at small $\gamma \ll 1$):
\begin{align}\label{expanu}
&r^{2(d-1)}\Big(r^2g_{tt}^{(0)}+g_{tt}^{(2)}+\frac{g_{tt}^{(4)}}{r^2}+...\Big)\Big(1+2\gamma \mathcal{C}^2\Big)=1.
\end{align}
If one writes the bulk Weyl squared tensor in terms of the boundary coordinates:
\begin{align}
\mathcal{C}^2=\mathcal{C}_{\mu \nu \rho \sigma}\mathcal{C}^{\mu \nu \rho \sigma}=\mathcal{C}_{ijkl}\mathcal{C}^{ijkl}+4\mathcal{C}_{i\rho j\rho}\mathcal{C}^{i\rho j\rho}+4\mathcal{C}_{ijk\rho}\mathcal{C}^{ijk\rho}+...,
\end{align}
where $i, j, k, l$ are boundary coordinates and $\rho=\frac{1}{r}$ the radial. Using the Fefferman-Graham coordinate to write the bulk Weyl squared tensor in terms of the boundary values it is given by (see Appendix \ref{weylsquared}) \footnote{It should be noted that we used the FG coordinates at the finite cut-off $r=r_j$ where $r_j$ is a location of a local maxima of the potential. In addition, \eqref{weyldecom} is the Wey squared tensor in $d+1=4+1$ dimensions.}:

\begin{align}
&\mathcal{C}^2=\rho^4 \mathscr{C}^2+\rho^4Log\rho^2\mathcal{A}+\rho^6\mathcal{B}+\rho^8\Big((\nabla_ig^{(4)}_{jk})^2+g^{(4)}_{ij}g^{(4)ij}\Big)+...,\label{weyldecom}\\
&=\rho^4\mathcal{C}_{(4)}^2+\mathcal{O}(\rho^6),
\end{align}
where $\mathscr{C}^2$ is the boundary Weyl tensor squared, $\mathcal{A}$ and $\mathcal{B}$ are constant \cite{holrecon}. \eqref{weyldecom} is a pure geometrical expansion of the bulk Weyl squared tensor. Expressing \eqref{expanu} in the FG coordinates at order $\rho^2$ yields:
\begin{align}
\rho^2 g^{(4)}_{tt}+2\gamma \frac{g^{(0)}_{tt}}{\rho^2} \Bigg(\rho^4 \mathscr{C}^2+\rho^4Log\rho^2\mathcal{A}+\mathcal{O}(\rho^{4})\Bigg)=1.\label{jump}
\end{align}
\eqref{jump} shows that the jumps in the complexity are governed by dynamics of the energy-momentum tensor of the boundary \footnote{The expectation value of the boundary energy-momentum tensor is computed by the Brown-York Lagrangian (BY) in the bulk $<T_{ij}>=\frac{\partial S_{ren.}}{\partial g^{(0)}_{ij}}$. BY term expansion contains $g^{(4)}_{ij}$ \cite{sken}. In addition, the Weyl anomaly of the boundary theory is computed by the variation of the bulk Lagrangian under the conformal transformation. The density of the holographic Weyl anomaly comes from the coefficient of logarithm divergence \cite{sken}. Moreover, the Weyl anomaly in the boundary theory comes from the correlation of the boundary energy-momentum tensor. All in all, the bulk Weyl squared tensor is governed by the boundary energy-momentum tensor.}.

If there is non-vanishing Weyl squared tensor at the boundary, then it is expected that the jumps in the complexity at the finite cut-off $\rho=\rho_i$ are related to \footnote{\eqref{divener} is written in general dimension, where the logarithmic term is of order $d$.}:
\begin{align}
\mathscr{C}^2=\frac{\rho_i^{-2}-g^{(4)}_{tt}}{2\gamma g^{(0)}_{tt}}.\label{divener}
\end{align}
The above relation raises the question of whether the distance between the jumps corresponds to a physical quantity. The distance between the jumps could be realized by $d_{jumps}=|r_i-r_j|$ where $r_i$ means the location where the $"i"th $ jump occurs at. As it has been studied, the location of the jump is where the potential $\tilde{U}$ has a local maxima. In other words, roots of \eqref{cte}. The distance between roots depends on the quantities in \eqref{cte}, quantities in $g_{tt}$ and $\gamma$ the coefficient of the generalization. As an example we show that how jumps are controlled by the parameter of the generalization by \eqref{cte} \footnote{\eqref{partiald} is an approximation. The full statement is derived by the derivative of \eqref{cte}: $$\frac{\partial r}{\partial \gamma}\Big(\frac{\partial g_t}{\partial r}a^2 r^{2(d-1)}+2a \gamma g_t r^{2(d-1)} \frac{\partial \mathcal{C}^2}{\partial r}+2(d-1)g_t a^2r^{2d-3}\Big)=-2ag_tr^{2(d-1)}\mathcal{C}^2.$$ At large $r_f\gg1$ and in spherical symmetric black holes $\frac{\partial g_t}{\partial r}\& \frac{\partial \mathcal{C}^2}{\partial r}\ll \frac{1}{r}$. Then, we could neglect the first two terms in the R.H.S of the above equation.}:

\begin{align}\label{partiald}
\frac{\partial r_i}{\partial \gamma}=-\frac{r\mathcal{C}^2}{(d-1)}\Big|_{r=r_i}.
\end{align}
As a consequence the distance between jumps could be written:
\begin{align}
\frac{\partial \big(r_i-r_j\big)}{\partial \gamma}=\frac{1}{1-d} \Bigg(r\mathcal{C}^2\Big|_{r=r_i}-r\mathcal{C}^2\Big|_{r=r_j}\Bigg), \label{cutoff}
\end{align}
which means that the distance between jumps is encoded in the RG flow of the boundary theory i.e. the dynamics of the energy-momentum tensor of the boundary at different energy scales. Moreover, \eqref{cutoff} tells about the scale-dependent origin of the distance of the jumps in CGR.

Before proceeding, it is important to clarify the assumptions underlying our analysis and their domain of validity. First, we work in the limit of small and large \(\gamma\) (the parameter controlling the deviation from the standard complexity=volume proposal). Second, we employ the Fefferman–Graham expansion at a finite radial cutoff \(r_f\); this is a standard tool in holographic renormalization but assumes that the bulk geometry is asymptotically AdS and that the cutoff is sufficiently large (near the boundary). Third, we restrict to spherical symmetry for simplicity; extending to non-spherical backgrounds would introduce additional technical complications but is not expected to change the qualitative picture of branch competition. Fourth, the derivation of \ref{partiald} uses a large-\(r_f\) approximation, which is justified in the UV regime of the boundary theory. Within these approximations, the existence of competing branches and the associated critical points is robust. However, the precise location of the jumps in the CGR and their detailed dependence on the boundary energy-momentum tensor is more heuristic and may receive corrections beyond the perturbative regime. We therefore treat the CSL equation derived below as an effective description near criticality, rather than an exact non-perturbative statement.
\section{Critical behavior and a Callan-Symanzik-like equation}\label{cr}
As shown above, the locations of these jumps are scale dependent. This result suggests that the CGR itself could behave differently in various scales. In this section we are going to study the behavior of CGR in different scales. In this manner CGR is a function of the generalization parameter; however, the location of these jumps $r_i$ is a function of the parameter, so we could use a chain rule and \eqref{partiald}:

\begin{align}
\frac{d \dot{\mathcal{C}}_{Any}}{d\gamma}\Big|_{r_i}=&\frac{\partial \dot{\mathcal{C}}_{Any}}{\partial \gamma}\Big|_{r_i}+\frac{\partial \dot{\mathcal{C}}_{Any}}{\partial r}\frac{\partial r}{\partial \gamma}\Big|_{r_i}\\
=&\frac{\partial \dot{\mathcal{C}}_{Any}}{\partial \gamma}\Big|_{r_i}+\frac{r\mathcal{C}^2}{(1-d)}\frac{\partial \dot{\mathcal{C}}_{Any}}{\partial r}\Big|_{r_i}.\label{cs}
\end{align}
One could proceed with the leading order:
\begin{align}\label{chainr}
\frac{d \dot{\mathcal{C}}_{Any}}{d\gamma}\Big|_{r_i}=\frac{\partial \dot{\mathcal{C}}_{Any}}{\partial \gamma}\Big|_{r_i}+\frac{\mathscr{C}^2}{(1-d)r^3_i}\frac{\partial \dot{\mathcal{C}}_{Any}}{\partial r}\Big|_{r_i}.
\end{align}
Around $r_i$, where it is a late time and asymptotic behavior of $\dot{\mathcal{C}}_{Any}$ in Fefferman-Graham coordinates at $r_i$ cut-off, CGR at the leading order could be written as:
\begin{align}
&\dot{\mathcal{C}}_{Any}\sim P_v=\sqrt{\tilde{U}}=\sqrt{g_{tt}a^2r^{2(d-1)}}\\
&=\Bigg(r_i^{2(d-1)}\Big(r_i^2g_{tt}^{(0)}+g_{tt}^{(2)}+\frac{g_{tt}^{(4)}}{r_i^2}+...\Big)\Big(1+\gamma (\frac{\mathcal{C}_{(4)}^2}{r_i^4}+\frac{\mathcal{C}_{(6)}^2}{r_i^6}+...)\Big)^2\Bigg)^{\frac{1}{2}}\\
&= \Big(r_i^{2d}g_{tt}^{(0)}+r_i^{2(d-1)}g_{tt}^{(2)}+r_i^{2(d-2)}(g_{tt}^{(4)}+2\gamma g_{tt}^{(0)}\mathcal{C}_{(4)}^2)+r_i^{2(d-3)}(g_{tt}^{(5)}+2\gamma g_{tt}^{(4)}\mathcal{C}_{(4)}^2)\nonumber \\
&+r_i^{2(d-4)}(\gamma^2 g_{tt}^{(0)}\mathcal{C}_{(4)}^2\mathcal{C}_{(4)}^2)+...\Big)^{\frac{1}{2}}\label{cor}. 
\end{align} 
As a consequence, if $\gamma \rightarrow \lambda^{\nu} \gamma$ and $r_i \rightarrow \lambda^{\mu}r_i$, then $\dot{\mathcal{C}}_{Any}(\lambda^{\mu}r_i,\lambda^{\nu} \gamma)= \lambda^{\Delta}\dot{\mathcal{C}}_{Any}(r_i,\gamma)$, where $\Delta$, $\mu$ and $\nu$ are critical exponents. Derivative of CGR with respect to $\lambda$ and setting $\lambda=1$ is given:
\begin{align}
\frac{d\dot{\mathcal{C}}_{Any}}{d\lambda}\Big|_{\lambda=1}&=\frac{\partial \dot{\mathcal{C}}_{Any}}{\partial r_i}\frac{\partial r_i}{\partial \lambda}+\frac{\partial \dot{\mathcal{C}}_{Any}}{\partial \gamma}\frac{\partial \gamma}{\partial \lambda}\\
&=\frac{\partial \dot{\mathcal{C}}_{Any}}{\partial r_i}(\mu r_i)+\frac{\partial \dot{\mathcal{C}}_{Any}}{\partial \gamma}(\nu \gamma). \label{critical}
\end{align}
In addition:
\begin{align}
\frac{d\dot{\mathcal{C}}_{Any}(\lambda^{\mu}r_i, \lambda^{\nu} \gamma)}{d\lambda}\Big|_{\lambda=1}=\frac{d}{d\lambda}\Big(\lambda^{\Delta}\dot{\mathcal{C}}_{Any}(r_i, \gamma)\Big)\Big|_{\lambda=1}=\Delta \dot{\mathcal{C}}_{Any}(r_i, \gamma). \label{delta}
\end{align}
Combining \eqref{critical} and \eqref{delta}:
\begin{align}
\Big((\frac{\nu \gamma\mathscr{C}^2}{(1-d)r_i^3}-\mu r_i)\frac{\partial }{\partial r_i}-\nu \gamma \frac{d }{d\gamma}+\Delta \Big) \dot{\mathcal{C}}_{Any}=0, \label{cs}
\end{align}
where $(\frac{\nu \gamma\mathscr{C}^2}{(1-d)r_i^5}-\mu r_i)\frac{\partial }{\partial r_i}$ shows how CGR is changed by the scale of the energy $r_i$, $-\nu \gamma \frac{d \dot{\mathcal{C}}}{d\gamma}$ indicating how CGR is changed by the generalization parameter, in other words, if the CGR is increasing by increase in $\gamma$ or decreasing. We call \eqref{cs} \textit{Callan-Symanzik-like} equation in information theory, where $r_i$ is the information renormalization scale, $\beta=-\nu \gamma$ beta function in information theory and $\Delta$ information anomalous dimension.

\eqref{cor} suggests that CGR at $\gamma \ll 1$, near $r_i$ is scaled:
\begin{align}
&\dot{\mathcal{C}}_{Any}(\lambda^{\mu}r_i, \lambda^{\nu}\gamma)\sim  \Big(\lambda^{2d\mu}r_i^{2d}g_{tt}^{(0)}+\lambda^{2(d-1)\mu}r_i^{2(d-1)}g_{tt}^{(2)}+\lambda^{2(d-2)\mu}r_i^{2(d-2)}g_{tt}^{(4)}\nonumber \\
&+\lambda^{2(d-2)\mu+\nu}r_i^{2(d-2)}2\gamma g_{tt}^{(0)}\mathcal{C}_{(4)}^2+\lambda^{2(d-3)\mu}r_i^{2(d-3)}g_{tt}^{(5)}+\lambda^{2(d-3)\mu+\nu}r_i^{2(d-3)}2\gamma g_{tt}^{(4)}\mathcal{C}_{(4)}^2\Big)^{\frac{1}{2}}\\
&=\lambda^{\Delta}\dot{\mathcal{C}}_{Any}(r_i,\gamma)
\end{align}
Now we could consider the validity of critical exponents for arbitrary dimensions:
\begin{align}
&\Delta=\frac{\nu}{2},\label{small}\\
&\mu=0.
\end{align}
However, for large $\gamma \gg 1$ we could write $a^2\approx 1+\gamma^2 \mathcal{C}^4$. Then \eqref{cor} is scaled as:
\begin{align}
&\dot{\mathcal{C}}_{Any}\sim P_v=\sqrt{g_{tt}a^2r^{2(d-1)}}=\Big(\lambda^{2d\mu}r_i^{2d}g_{tt}^{(0)}+\lambda^{2(d-1)\mu}r_i^{2(d-1)}g_{tt}^{(2)}+\lambda^{2(d-2)\mu}r_i^{2(d-2)}g_{tt}^{(4)}\nonumber \\
&+\lambda^{2(d-3)\mu}r_i^{2(d-3)}g_{tt}^{(5)}+\lambda^{2(d-4)\mu+2\nu}r_i^{2(d-4)}\gamma^2 g_{tt}^{(0)}\mathcal{C}_{(4)}^2\mathcal{C}_{(4)}^2\Big)^{\frac{1}{2}}.
\end{align} 
The critical exponents for arbitrary dimensions would be:
\begin{align}
&\Delta=\nu,\label{large}\\
&\mu=0.
\end{align}
As \eqref{small} and \eqref{large} show, the scaling of CGR is different at different regimes of the generalization parameter. Looking again at \eqref{cs} by the solution reveals:

\begin{align}
\Big((\frac{\nu \gamma\mathscr{C}^2}{(1-d)r_i^3})\frac{\partial }{\partial r_i}-\nu \gamma \frac{d }{d\gamma}+\Delta \Big) \dot{\mathcal{C}}_{Any}=0.
\end{align}
The solution of the above equation would be:
\begin{align}
\dot{\mathcal{C}}_{Any}=\gamma^{n}\mathcal{F}(r^6)+\alpha,
\end{align}
where $\mathcal{F}$ is a function of $r^6$, $n=1$ at $\gamma \gg1$, $n=\frac{1}{2}$ at $\gamma \ll1$ and $\alpha$ is the constant in \eqref{cte}.
By the fact that $\dot{\mathcal{C}}$ is constant at $r_i$, then it could be written as:
\begin{align}
\dot{\mathcal{C}}_{Any}=\gamma^n \mathcal{F}(r^6-r_i^6(\gamma)+\alpha.
\end{align}
This indicates how the CGR scales with the correlation length $|r - r_i|$ near the critical point for any arbitrary dimension.

\section{Conclusion}\label{con}
The AdS/CFT correspondence, by relating a gravitational theory to a quantum field theory, can alleviate the difficulties of perturbation theory in strongly-coupled systems. Exploring the counterpart quantities in the AdS/CFT make it a strong technique to study the physical theories. Computational complexity, as one of these quantities, has been extensively studied in holography through several proposals. Recently, a proposal known as complexity equals anything has been introduced where the complexity is identified with the volume of a generalized hypersurface. Studying the complexity of various models through this proposal shows some branches in CGR. Moreover, these branches has been considered as a phase transition in CGR.

Since phase transitions appear in the "complexity = anything" framework, this suggests that the generalized surfaces can undergo the transition. One mechanism for the generalization is the Weyl squared tensor. Expanding the Weyl squared tensor in the Fefferman-Graham coordinate makes it to be reduced into two terms: divergence of the energy-momentum tensor and the Weyl squared tensor of the boundary which is only a geometrical expansion. However, the phase transitions and jump separations are governed by the boundary energy-momentum tensor.

CGR in this way behaves as a function of the generalization parameter and the locations of the jumps. In other words, these two parameters control CGR. Considering the behavior of CGR around the phase transition reveals a scaling behavior and critical exponents. This shows that interpreting these jumps as phase transitions is meaningful. However, CGR satisfies a Callan-Symanzik-like (CSL) equation in field theory in which a beta function is defined in terms of the generalization parameter and an anomalous dimension in terms of the scaling of CGR, which means that the critical behavior of CGR is universal regardless of the dimension.

A recent work in phase transitions and "complexity=anything" in low-dimensional dilaton gravities \cite{2503} interprets them in two ways. First, each maximum represents a distinct locally optimal quantum circuit that prepares the same final state using different gate sets, with the globally optimal circuit giving the true complexity via a minimax prescription. Second, all local maxima contribute additively to the total complexity when the system factorizes into independent, disentangled subsectors, each with its own optimal path. The CSL equation naturally incorporates these transitions and aligns well with both interpretations. The first interpretation suggests that there are alternative quantum circuits in which one of them is dominated and the computation finds different gate set and carried out with the optimal one, in other words, the CSL equation identifies which branch is locally optimized. In the second interpretation, the circuit is assumed to contain several disentangled subsectors that independently carry out the computation, with each maximum in the effective potential representing the optimal path in a given subsector. The total complexity is then the sum of the contributions from all subsectors. The CSL equation, in this interpretation, captures the overall performance of all subsectors. Thus, the transitions governed by the CSL equation depict a transition from one subsector to another i.e., the optimal path in each subsector is represented by \(\mathcal{C}_i\) (the contribution from the \(i\)-th local maximum), and the total complexity is \(\mathcal{C} = \sum_i \alpha_i \mathcal{C}_i\).

Finally, we briefly comment on the possible broader implications. The observation that the CGR satisfies a CSL equation suggests an interesting parallel between holographic complexity and the renormalization group. In principle, this could hint at a scenario where tuning the energy scale in a quantum circuit (via a scale-dependent gate set) leads to qualitatively different rates of information processing. However, we must emphasize that the "complexity = anything'' proposal is a holographic conjecture whose precise relation to standard definitions of quantum circuit complexity (or to actual quantum computers) is still far from established. Our results are best viewed as a theoretical exploration of the structure of holographic phase transitions; any connection to real-world computational devices remains highly speculative at present. We hope that future work, possibly along the lines of \cite{2503}, will help bridge this gap.

\section*{Acknowledgment}
The authors would like to express their gratitude to Monireh Emami for useful discussions. In addition, OpenAI's ChatGPT has been used to enhance the clarity of this manuscript. 

\appendix
\renewcommand\theequation{\thesection-\arabic{equation}} 
\setcounter{equation}{0}

\section{Relation between turning time and the maximum of the effective potential}\label{potjump}
In this appendix we clarify the relation between the turning time of the generalized complexity and the heights of the local maxima of the effective potential.

Consider an effective potential $\tilde{U}(r)$ governing extremal hypersurfaces for generalized complexity. Suppose that $\tilde{U}(r)$ admits two local maxima at:
\[
r = r_{f,1}, \qquad r = r_{f,2},
\]
with corresponding heights:
\[
\tilde{U}_{f,i} \equiv \tilde{U}(r_{f,i}), \qquad i=1,2.
\]

Each local maximum defines a branch of extremal hypersurfaces \cite{jump}. The conserved momentum $P_v$ on each branch satisfies:
\[
P_v^2 \to \tilde{U}_{f,i}
\qquad
(\tau \to \infty),
\]
and the complexity on each branch exhibits linear late-time growth of the form:
\begin{equation}
\mathcal{C}_{Any}^i(\tau)
=
\sqrt{\tilde{U}_{f,i}}\,\tau + \beta_i + \mathcal{O}(1),
\label{eq:late_time_complexity}
\end{equation}
where $\beta_i$ are branch-dependent constants.

The late time growth rate is therefore determined solely by the height of the corresponding maximum of the effective potential.

\subsection{Turning time and dependence on potential heights}
The turning time $\tau_{\mathrm{turn}}$ is defined as the boundary time at which the dominant extremal hypersurface switches between the two branches:
\begin{equation}
\mathcal{C}_{Any}^1(\tau_{\mathrm{turn}})
=
\mathcal{C}_{Any}^2(\tau_{\mathrm{turn}}).
\label{eq:turning_def}
\end{equation}

Using the late-time expansion \eqref{eq:late_time_complexity}, one finds:
\begin{equation}
\tau_{\mathrm{turn}}
=
\frac{\beta_2-\beta_1}
{\sqrt{\tilde{U}_{f,1}}-\sqrt{\tilde{U}_{f,2}}}
+ \mathcal{O}(1),
\label{eq:turning_time_formula}
\end{equation}
provided $\sqrt{\tilde{U}_{f,1}} \neq \sqrt{\tilde{U}_{f,2}}$.

\eqref{eq:turning_time_formula} shows explicitly that the turning time is controlled by the difference in the heights of the effective potential maxima. In particular:
\begin{equation}
|\sqrt{\tilde{U}_{f,1}}-\sqrt{\tilde{U}_{f,2}}| \ll 1
\quad \Longrightarrow \quad
\tau_{\mathrm{turn}} \gg 1.
\end{equation}

\subsection{Relation to the radial position of the transition}

Independently of branch switching, the boundary time diverges logarithmically as the turning point $r_{\min}$ approaches a local maximum $r_f$ of the effective potential \cite{jump}:
\begin{equation}
\tau \sim
\frac{1}{\sqrt{|\tilde{U}''(r_f)|}}
\ln\!\left(\frac{1}{\tilde{U}(r_f)-P_v^2}\right).
\end{equation}
Consequently, at parametrically large boundary time:
\begin{equation}
r_{\min}(\tau)
=
r_f + \mathcal{O}\!\left(e^{-\kappa \tau}\right),
\end{equation}
with $\kappa>0$.

Combining this result with \eqref{eq:turning_time_formula}, we conclude that when the potential maxima are nearly degenerate in height, the turning time occurs at sufficiently large $\tau$ such that the competing extremal hypersurfaces probe regions exponentially close to the corresponding maxima. In this regime, the transition in the complexity growth rate appears to occur around $r_f$.

\subsection{Why near-degenerate maxima}

\eqref{eq:turning_time_formula} explains why most examples studied in the literature focus on cases where:
\[
\tilde{U}_{f,1} \simeq \tilde{U}_{f,2}.
\]
If the heights of the maxima differ by an $\mathcal{O}(1)$ amount, the turning time is also $\mathcal{O}(1)$ and the branch transition occurs before the late-time regime is reached. In such cases, the extremal hypersurfaces involved in the transition do not probe the vicinity of the potential maxima, and the transition is neither sharp nor universal.

By contrast, near-degenerate maxima lead to a parametrically large turning time, a sharp transition in the CGR, and a clean geometric interpretation in terms of the near-maximum structure of the effective potential. For this reason, examples with nearly equal potential heights are naturally emphasized in the literature and considered.

The turning time of the generalized complexity is controlled by the difference in the heights of competing local maxima of the effective potential. Near-degeneracy of these maxima leads to a late-time transition governed by extremal hypersurfaces probing the vicinity of the maxima, leads to sharp transitions.

\section{Weyl squared tensor in FG coordinate}\label{weylsquared}
In this appendix, we derive the expansion of the bulk Weyl squared tensor $\mathcal{C}^2 = \mathcal{C}_{\mu\nu\rho\sigma}\mathcal{C}^{\mu\nu\rho\sigma}$ in Fefferman-Graham (FG) coordinates, which is used in the main text \eqref{weyldecom} to relate the jump locations in the complexity growth rate to the boundary energy-momentum tensor.

\subsection{Fefferman-Graham expansion}

The FG coordinate system is defined by the line element:
\begin{align}
&ds^2=G_{\mu\nu}dx^{\mu}dx^{\nu}=\frac{l^2}{\rho^2}\Big(d\rho^2+g_{ij}(x,\rho)dx^idx^j\Big),
\end{align}
where $\rho = 0$ corresponds to the asymptotic AdS boundary. The induced metric $g_{ij}(x,\rho)$ admits the asymptotic expansion:
\begin{align}
g_{ij}(x,\rho) = g^{(0)}_{ij}(x) + \rho^2 g^{(2)}_{ij}(x) + \rho^4 g^{(4)}_{ij}(x) + \cdots + \rho^{d}\big(g^{(d)}_{ij}(x) + h^{(d)}_{ij}(x)\log\rho^2\big) + \cdots,
\end{align}
valid for asymptotically AdS spacetimes in $d+1$ dimensions. In the main text, we work in $d=4$ (five bulk dimensions) for simplicity, where the expansion simplifies to:
\begin{align}
g_{ij}(x,\rho) = g^{(0)}_{ij}(x) + \rho^2 g^{(2)}_{ij}(x) + \rho^4\left(g^{(4)}_{ij}(x) + h^{(4)}_{ij}(x)\log\rho^2\right) + \mathcal{O}(\rho^6),
\end{align} 
where the logarithmic term $h^{(4)}_{ij}$ appears when the boundary dimension is even and encodes the Weyl anomaly and $g^{(2)}_{ij}=-P_{ij}$ the Schouten tensor of $g^{(0)}_{ij}$ and $g^{(4)}_{ij}\sim <T_{ij}>$ in five dimensions \cite{sken}:
\begin{align}
P_{ij}=\frac{1}{2}\Big(R_{ij}-\frac{Rg^{(0)}_{ij}}{6}\Big).
\end{align}

\subsection{Connection and curvature components}
From the FG ansatz, the non-vanishing Christoffel symbols are:
\begin{align}
&\Gamma_{\rho \rho}^{\rho}=-\frac{1}{\rho},\\
&\Gamma_{ij}^{\rho}=-\frac{1}{\rho}g_{ij}+\frac{1}{2}\partial_{\rho}g_{ij}=2\rho g^{(2)}_{ij}+4\rho^3(g^{(4)}_{ij}+h^{(4)}_{ij}Log\rho^2)+...,\\
&\Gamma_{\rho j}^i=-\frac{1}{\rho}\delta_j^i+\frac{1}{2}g^{ik}\partial_{\rho}g_{kj}.
\end{align}
Using these, the Riemann tensor, the Ricci tensor components and the Ricci scalar relevant for the Weyl tensor are:
\begin{align}
&R_{ijkl}(G)=R_{ijkl}(g|_{\rho})-\frac{1}{\rho^2}(g_{ik}g_{jl}-g_{il}g_{jk})+\frac{1}{4}\partial_{\rho}g_{ik}\partial_{\rho}g_{jl}-\frac{1}{4}\partial_{\rho}g_{il}g_{jk}+...,\\
&R_{\rho i \rho j}(G)=-\frac{1}{2}\partial_{\rho}^2g_{ij}+\frac{1}{4}g^{kl}\partial_{\rho}g_{ik}\partial_{\rho}g_{jl}=-\frac{1}{\rho^2}g^{(0)}_{ij}+... ,\\
&R_{\rho ijk}(G)=\frac{1}{2}(\nabla_j \partial_{\rho}g_{ik}-\nabla_k \partial_{\rho}g_{ij})=\nabla_j (\rho g^{(2)}_{ik})-\nabla_k (\rho g^{(2)}_{ij}),\\
&R_{\rho \rho}=-\frac{1}{2}g^{ij}\partial_{\rho}^2g_{ij}+\frac{1}{4}g^{ik}g^{jl}\partial_{\rho}g_{ij}\partial_{\rho}g_{kl},\\
&R_{\rho i}=\frac{1}{2}g^{jk}(\nabla_k \partial_{\rho}g_{ij}-\nabla_i \partial_{\rho}g_{jk}),\\
&R_{ij}=R_{ij}(g|_{\rho})-\frac{1}{2}\partial_{\rho}^2g_{ij}+\frac{1}{4}g^{kl}\partial_{\rho}g_{ik}\partial_{\rho}g_{jl}+...=R_{ij}(g|_{\rho})-4g^{(2)}_{ij}+...,\\
&R=-\frac{20}{l^2}+\rho^2R(g^{(0)})+...,
\end{align}
where $g|_{\rho}=g^{(0)}_{ij}+\rho^2g^{(2)}_{ij}+...$ and $\nabla$ is with respect to $g|_{\rho}$.

\subsection{Expansion of the Weyl tensor}
The Weyl tensor in $d+1$ dimensions is defined as:
\begin{align}
\mathcal{C}_{\mu\nu\rho\sigma} = R_{\mu\nu\rho\sigma} - \frac{1}{d-1}\big(g_{\mu\rho}R_{\nu\sigma} - g_{\mu\sigma}R_{\nu\rho} - g_{\nu\rho}R_{\mu\sigma} + g_{\nu\sigma}R_{\mu\rho}\big) + \frac{1}{d(d-1)}\big(g_{\mu\rho}g_{\nu\sigma} - g_{\mu\sigma}g_{\nu\rho}\big)R.
\end{align}
Substituting the expansions of the metric and curvature, and keeping terms up to $\mathcal{O}(\rho^8)$ in the square of the Weyl tensor, one obtains:
The Weyl tensor is given by:
\begin{align}
\mathcal{C}^2=\rho^4 \mathscr{C}+\rho^4Log\rho^2\mathcal{A}+\rho^6\mathcal{B}+\rho^8\Big((\nabla_ig^{(4)}_{jk})^2+g^{(4)}_{ij}g^{(4)ij}\Big)+...,
\end{align}
where:
\begin{itemize}
\item $\mathscr{C}^2(x)$ is the square of the boundary Weyl tensor (constructed from $g^{(0)}_{ij}$),
\item $\mathcal{A}$ and $\mathcal{B}$ are local expressions involving $g^{(0)}_{ij}$, $g^{(2)}_{ij}$, and their derivatives,
\item The $\rho^8$ term involves the boundary energy-momentum tensor through $g^{(4)}_{ij}$.
\end{itemize}

\end{document}